# Erosion rate of lunar soil under a landing rocket, part 1: identifying the rate-limiting physics

**Authors:**

Philip T. Metzger[1]*

**Affiliations:**

[1] Stephen W. Hawking Center for Microgravity Research and Education, University of Central Florida, Orlando, FL 32828, USA.

*Corresponding author. Email: Philip.metzger@ucf.edu

Abstract: Multiple nations are planning activity on the Moon's surface, and to deconflict lunar operations we must understand the sandblasting damage from rocket exhaust blowing soil. Prior research disagreed over the scaling of the erosion rate, which determines the magnitude of the damage. Reduced gravity experiments and two other lines of evidence now indicate that the erosion rate scales with the kinetic energy flux at the bottom of the laminar sublayer of the gas. Because the rocket exhaust is so fast, eroded particles lifted higher in the boundary layer do not impact the surface for kilometers (if at all; some leave the Moon entirely), so there is no saltation in the vicinity of the gas. As a result, there is little transport of gas kinetic energy from higher in the boundary layer down to the surface, so the emission of soil into the gas is a surprisingly low energy process. In low lunar gravity, a dominant source of resistance to this small energy flux turns out to be the cohesive energy density of the lunar soil, which arises primarily from particles in the 0.3 to 3 $\mu$m size range. These particles constitute only a tiny fraction of the mass of lunar soil and have been largely ignored in most studies, so they are poorly characterized.

## 1. Introduction

When a spacecraft uses rocket propulsion to land on the Moon or another body, the exhaust gas accelerates particulates to high speed [1–3], threatening the spacecraft and nearby hardware [4] while disturbing mission science. A lunar lander the size of the Apollo Lunar Module (LM) sends ejecta globally since the exospheric atmosphere is too thin to slow even the finest dust. Some dust exceeds escape velocity and may damage lunar orbiters [5]. Parts of the physics have remained unsolved so we cannot predict the necessary size of no-landing zones around sensitive hardware or operations. NASA recommended a 2 km no-landing zone around the lunar heritage artifacts [6], but it was this author and a colleague who selected the 2 km distance arbitrarily because it seemed reasonable as a placeholder until we can learn more.[1] We discussed at the time that the document would be a "living document" to be updated after determining a realistic landing distance. It is not unlikely that sensitive hardware such as telescopes could require 10 km radius or larger unless other mitigation is established. Whatever policy is decided should be based in science, so this field of research is urgent.

---

[1] The decision was made by Philip Metzger (this author) and Rob Kelso, both of NASA at the time. We calculated 2 km as the distance, rounded to the nearest kilometer, to the horizon as seen by a 6 foot tall person. We discussed that the horizon is irrelevant to the physics because particles curve over it in gravity, but a 2 km limit seemed reasonable and was better than no limit since time was of the essence during the Google Lunar X-Prize. Note that the other parts of the document to protect the lunar heritage sites were based on a firmer scientific basis.

The relevant regime of fluid physics is understudied and differs from terrestrial wind-blown sand and dust because the gas is expanding into vacuum where the long mean free path lengths of the gas molecules suppress turbulence and alter the boundary layer, and it is supersonic, so saltation [7] plays no role as lifted particles do not fall for kilometers beyond the gas flow, if at all. In terrestrial erosion physics, the emission rate of dust is controlled by the saltation of sand transporting energy from the higher velocity gas down to the surface [8], so progress in that field is not applicable. The problem is compounded by the exotic nature of lunar soil with unweathered, jagged particle shapes and high dust content due to lack of geological sorting, with the effects of cohesion amplified by reduced gravity. It is difficult to study the physics because we cannot fire a sufficiently large rocket engine inside a vacuum chamber while maintaining vacuum, with abrasive dust that would damage vacuum pumps, with the entire experiment inside a reduced gravity aircraft. The problem has been approached through a variety of experiments to access different parts of the parameter space coupled with analysis of videos taken during lunar landings and computer simulations that incorporate the parts of the physics that we understand.

This effort has identified several distinct regimes in which gas jets move the soil. One of these has been called *viscous erosion* [9], surface erosion, or simply erosion. It is the process where gas flows across the soil's surface, lifting grains and accelerating them away. The other regimes are collectively called *deep cratering* because they move soil in bulk with the potential to quickly form deep craters. These regimes include bearing capacity failure [10], diffused gas eruption [11], diffusion driven shearing [12], and diffused gas explosive erosion [13]. The deep cratering mechanisms generally did not occur during lunar landings due to the high mechanical strength and low permeability of the soil and because rocket exhaust in vacuum is not collimated into a jet, so it produces smaller gradients of pressure on the surface. However, during the final moments of the Apollo landings when the engine nozzle was close to the surface, an abrupt blast of soil sometimes occurred, sending out bulk masses or clods that were fragmenting as they blew away [14]. These events may have involved bearing capacity failure or another deep cratering mechanism, although they removed only a shallow layer of soil, resulting in a flat surface under the lander. The plume and soil conditions where deep cratering will occur are not yet known, and it is possible that larger lunar landers could dig deep craters.

## 2. The Unsolved Erosion Rate Scaling

Better progress has been made explaining surface erosion, which is the primary mechanism during lunar landings and is the focus of this paper. Some of the scaling for the surface erosion rate has been identified in experiments by Haehnel [15], Donahue et al. [16,17], Metzger et al. [12,18,19], LaMarche et al. [20], and Metzger [21], but the theory was incomplete so could not be extrapolated reliably to the lunar environment.

Researchers disagreed whether the erosion rate should be proportional to $\tau$ = shear stress in the gas, $\rho v^2$ = momentum flux (where $\rho$ = gas density and $v$ = gas velocity), $\rho v^3/2$ = kinetic energy flux, or $\tau^{2.5}$ as suggested by one analysis of the Apollo lunar landings [1,18,22–24]. The rate has been studied in small-scale experiments where a jet of gas impinges vertically on a granular material [12,15–21,25–28]. These experiments are useful because erosion occurs over a limited distance, so saltation plays no role, and because the size and shape of the portion of the crater where gas interacts with the granular material are naturally constant over a long time (discussed below) enabling measurement in steady flow conditions. The small scale of these experiments enables performing them in vacuum chambers and reduced gravity aircraft.

These experiments typically varied one parameter at a time and found the power index on each parameter that matches how the erosion rate changes. Metzger et al. [18] combined these results and argued that to form a dimensionless group the scaling should be,

$$\dot{V} = \frac{\dot{M}}{\rho_b} = C \frac{\rho_e v_e^2 A_e}{\rho_b g \langle D \rangle} \tag{1}$$

where $\dot{V}$ = volumetric erosion rate (m³/s), $\dot{M}$ = mass erosion rate (kg/s), $\rho_b$ = bulk density of the soil, $\rho_e$ = gas density at the nozzle exit, $v_e$ = gas velocity at the nozzle exit, $A_e$ = exit area of the nozzle, $g$ = gravity, and $\langle D \rangle$ = mean particle diameter of the soil grains. They noted the proportionality constant $C$ has units of velocity. It could be a molecular velocity (speed of sound in the gas) or an eddy transport rate since in the terrestrial experiments the gas jets were turbulent. This result was compelling because the numerator comprises properties of the gas that drive erosion while the denominator comprises properties of the soil that resist erosion. The denominator is in the form of a potential energy: the energy needed per volume of soil to lift it the tiny height $\langle D \rangle$, presumably the height each grain must be lifted to get over its neighboring grain. $C \rho_e v_e^2 A_e$ in the numerator has units of energy flux and may be interpreted as the rate the energy is delivered to perform the lifting process.

Reduced gravity experiments with a finer granular material were found by Metzger et al.[19] to depart from the 1/g scaling at the lowest gravity levels. They guessed it may be due to cohesion and that the effect was not seen with coarse sand even at lowest gravity due to sand's insignificant cohesion. Writing the denominator as the sum of these two energy densities,

$$\dot{V} = \frac{\dot{M}}{\rho_b} = C \frac{\rho_e v_e^2 A_e}{\rho_b g \beta \langle D \rangle + \alpha} \tag{2}$$

where $\alpha$ = cohesive energy per volume of soil (J/m³). $\beta = \cos\theta \approx 1/2$ has been included in the potential energy term because erosion took place on the inner slope of the craters where $\theta \approx 60°$ for tests in Earth's gravity.

Rajaratnam and Mazurek [28] showed for these turbulent experiments that local shear stress on the surface $\tau(\vec{R})$ is proportional to $\rho_e v_e^2 A_e$, where $\vec{R}$ = location on the surface. Substituting $\tau(\vec{R})$ in the numerator and multiplying by $\rho_b$ converts eq. 2 from total volumetric erosion of a scour hole to a local mass erosion rate $\dot{m}$ (in kg/m²/s),

$$\dot{m} = \rho_b \frac{C\,\tau(\vec{R})}{\rho_b g \beta \langle D \rangle + \alpha} \tag{3}$$

which can be integrated over the surface area to find the total mass erosion rate, $\dot{M} = \iint \dot{m}\,dA$ (in kg/s).

While the experiments showed $\langle D \rangle$ in eqs. 1 and 2 is proportional to $D_{50}$ = the median particle size of the material, it was not necessarily identical to $D_{50}$. For example, it could be some multiple of $D_{50}$ that particles must be lifted before they are swept downwind by the gas, or $0.7\,D_{50}$, which is relevant to laminar sublayer momentum transfer to the soil [29], or $D_{84}$ (the size such that 84% of the mass of the soil is finer), which is used in sediment transport studies [30]. For the baseline model of lunar soil used here, $D_{50} = 77.5\,\mu m$ and $D_{84} = 2.3\,D_{50}$. These values are mentioned here to provide a sense of scale. They are calculated in the companion paper, "Erosion rate of lunar soil under a landing rocket, part 2: benchmarking and predictions" (*this issue*), by integrating the model of the lunar soil particle size distribution. Interpreting $\langle D \rangle$ is crucial for determining where the rate-limiting process exists in the physics, such as energy or momentum consumption accelerating the entrained grains high in the boundary layer, or diffusion of energy across the laminar sublayer, or producing stagnation pressure beneath the grains that have not yet been lifted.

For example, Roberts [1] hypothesized that the rate of particles leaving the surface is determined by the acceleration of grains after they have been eroded, because that consumes momentum from the gas flow, which reduces shear stress on the surface. Roberts hypothesized that when shear stress of the gas has been reduced to equal the shear strength of the soil, then the erosion rate no longer increases. The particle size distribution determines the efficiency of momentum transfer from the gas to the particles, so his theory derived a parameter that has units of velocity, representing the asymptotic velocity the particles achieve, which is a function of the particle size. That velocity is analogous to $C$ in eq. 1. If Roberts' hypothesis is correct, then the rate limiting physics is high above the surface where the eroded grains are being accelerated. His hypothesis is still being used and applied to future lunar landings [23,31], so we need to know whether it is correct.

### 3. Evidence for Completing the Theory

It is difficult to make progress without new, carefully designed, expensive flight experiments both terrestrially in reduced gravity and on the Moon. Before the expense can be justified to a funding agency, we must develop and test the theory as far as possible without them. We have three lines of evidence that have bearing on the identities of $C$, $\langle D \rangle$, and $\alpha$. These lead to a hypothesis for completing the theory.

*3.1. Imagery of Eroded Soil*

The first line of evidence is from imagery of the eroded soil beneath the LMs and in related field tests. Metzger et al. [14] noted that the eroded craters had tiny, discrete "stair steps" of soil instead of smooth surfaces (fig. 1A). This was explained as geological micro-laminations in the soil column (impact ejecta blankets on the Moon, or seasonal deposition at terrestrial field sites) with a weak mechanical discontinuity at the interface of each layer. These fragile discontinuities had gone unnoticed in other soil mechanics measurements, but the rocket plume exposed the fine features like an archeologist's careful brushing to unearth an artifact. This suggests the removal of surface material without saltation is a low-energy process despite the violence of the plume above the boundary layer. Also, after each field test a fine layer of dust that flew away with the smallest mechanical disturbance was found coating the surface areas of the fresh erosion craters, suggesting that both erosion of soil and redeposition of only the finest particles were occurring simultaneously inside the crater (fig 1B). This surprising observation also suggests that erosion is a low-energy process. Gas flow deep in the laminar sublayer of the boundary is slow and thus low energy, so this suggests that transport of energy across the boundary to the soil is not efficient. This could be because it is not dominated by the energy transport of saltation, or of grains colliding and scattering back down from higher in the boundary, or of a thick bedload transport that is in contact with both the faster gas up high and the surface beneath (fig. 1C). Individual lifting of grains via stagnation pressure of the gas without these energetic particle dynamics, i.e., simply the lifting force from Bernoulli's principle, would be an example of a low energy erosion process (fig. 1D).

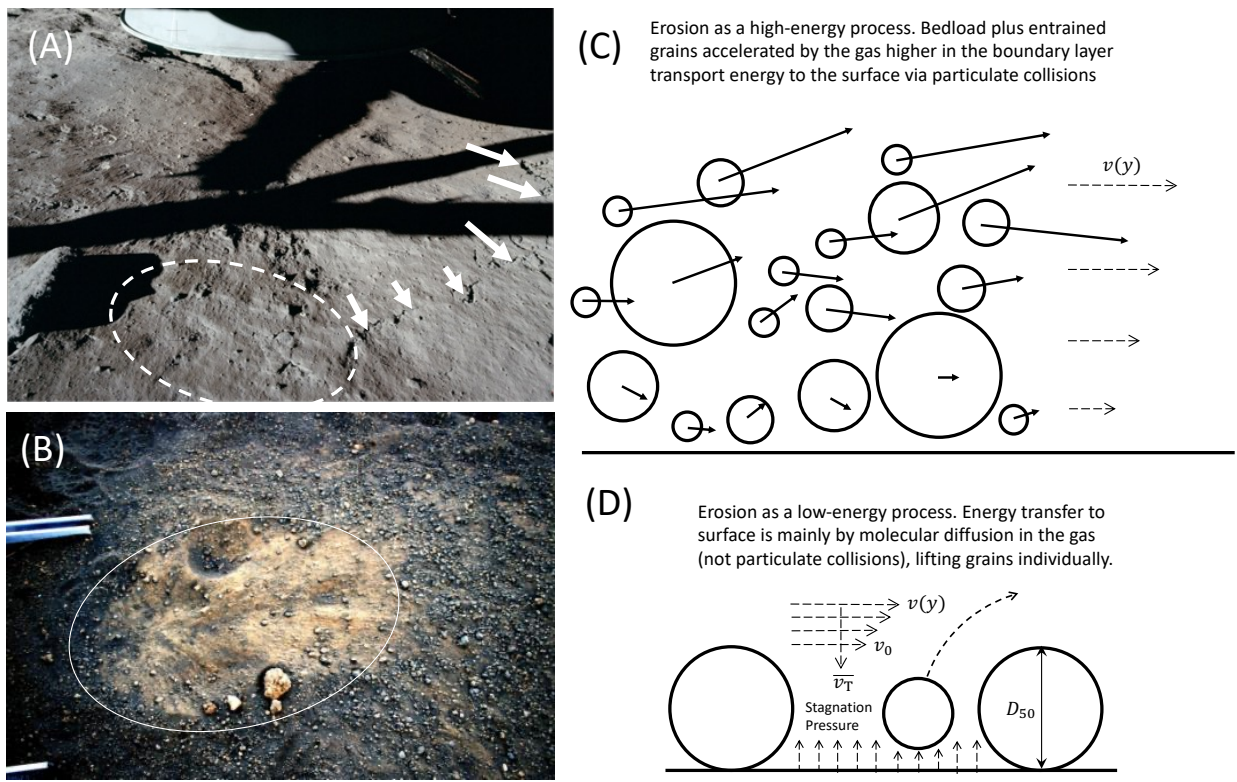

**Figure 1. High-energy versus low-energy erosion processes.** (A) Soil beneath the Apollo 11 engine nozzle with a sharp micro-scarp (arrows) and many rounded micro-scarps (ellipse) stepping downhill from the nozzle toward the radius of maximum erosion depth. (Image: NASA) (B) Erosion crater from a rocket plume on volcanic tephra, contrast enhanced. The bright dust veneer inside

the crater was deposited simultaneously with gas digging the crater. (C) Notional high-energy process. (D) Notional low-energy process where molecular thermal velocity $\overline{v_T}$ determines the rate of energy flux across the laminar flow.

*3.2. Functional Form of Crater Growth in Experiments*

The second evidence comes from the time evolution of the crater depth in the small-scale terrestrial experiments [12,15–20]. These found that when a jet is allowed to scour the surface for a sufficiently long time (usually only a fraction of a second, but sometimes much longer depending on test conditions), the crater will develop a compound shape as shown in fig. 2A with a parabolic inner crater and a conical outer crater. The gas flow from the jet interacts only with the inner crater as shown in fig. 2B, so the outer crater, with no external force other than gravity, relaxes to the angle of repose of the granular material. Therefore, lifting of grains from the surface by direct action of the gas takes place only in the inner crater, and primarily at the upper lip of the inner crater where shear stress of the gas is maximum. Eroded grains fly up and outside the gas flow, landing either in the outer crater or beyond, so there is no saltation in the region of erosion. This provides crucial similarity to the case of interest.

The upward traction of the gas along the upper rim of the inner crater also stabilizes the inner crater slope to be greater than the angle of repose. However, erosion deepens and widens the inner crater, which spreads the gas flow over a wider circumference at the lip, reducing the traction of the gas, so the upper part of the inner crater becomes unstable and avalanches. This removes the footing of the surface of the outer conical crater, so the surface of the entire outer crater avalanches in response, dumping particles into the gas flow inside the inner crater. Removing a layer from the outer crater by avalanche effectively lowers the lip of the inner crater, which restores the depth and width of the inner crater to the size that was stable. Thus, the inner crater remains essentially constant throughout the experiment as shown in fig. 2D. Paradoxically, all primary erosion occurs inside the inner crater, but all net crater growth is due to the increasing depth and width of the outer crater via avalanche. This has interesting consequences.

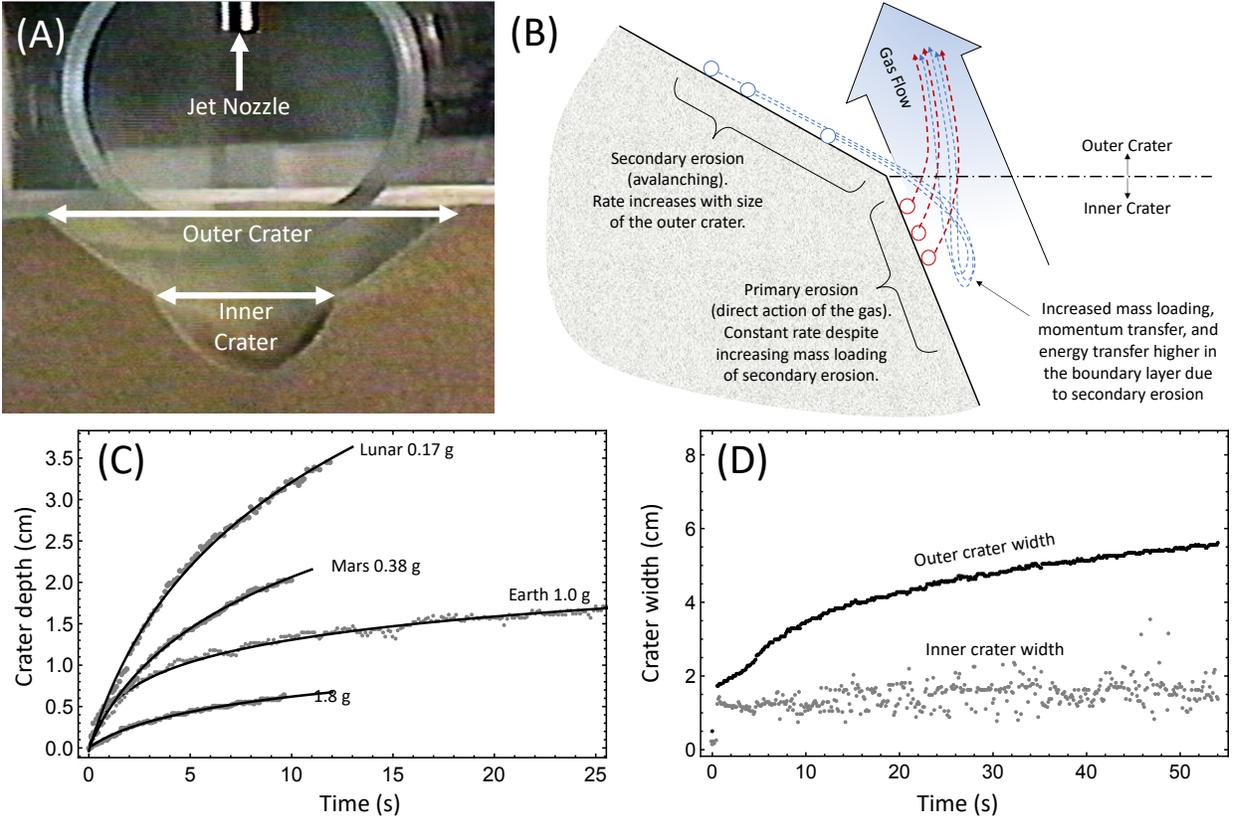

**Figure 2. Crater formation in small scale experiments.** (A) Crater morphology seen through plexiglass window. (B) Mass loading higher in the boundary layer due to secondary erosion. (C) Crater depth vs. time for different gravities fitted by logarithmic functions. (D) Inner and outer crater widths vs. time, showing the inner crater maintains constant size while the outer crater grows.

The craters are observed, as in fig. 2C, to grow in depth $H$ logarithmically over a significant period,

$$H = a \, \text{Log}(bt + 1) \tag{4}$$

where $a$ is a length scale and $1/b$ is a time scale. The depth eventually diverges from logarithmic growth and the crater size asymptotes after a typically long period varying with test conditions. The differential equation that produces a logarithmic form is

$$\frac{dH}{dt} = ab \, \text{Exp}(-H/a) \tag{5}$$

The reason an exponential appears in the governing physics is discussed further below. It must be related to the fraction of the ejected particles that fall back into the outer crater instead of beyond, so they are recirculated. As the crater deepens and widens recirculating an ever larger fraction of the ejecta, the rate of net crater growth is slowed. Because the distribution of ejected particle kinetic energies does not have an infinitely long tail like a Gibbs distribution, eventually the crater is so wide that no particle can escape and growth stops entirely, which may explain the eventual departure from logarithmic crater growth.

In any case, because the overall crater's volumetric growth is entirely in the outer crater, the overall growth rate follows the equation for a cone and can be written,

$$\frac{dV}{dt} = \frac{\pi}{3}\cot^2\varphi \frac{dH^3}{dt} = \pi \cot^2\varphi\, H^2 ab\, \mathrm{Exp}(-H/a)$$
(6)

where $\varphi$ = angle of repose. The geometric factors $\pi \cot^2\varphi\, H^2$ represent the surface area of the outer (conical) crater. $ab$ has units of m/s and represents the rate that this surface is being lowered through avalanche, so their product has units m³/s. The rate of avalanches $ab$ depends on the rate of erosion in the inner crater, which is therefore proportional to the constant value $ab$ throughout the experiment. The exponential factor $\mathrm{Exp}(-H/a)$ represents the fraction of ejecta that escape the outer cone and are not recycled. A simplistic argument can show why this form appears. We may approximate it as the Gibbs distribution of the ejecta's kinetic energies $E$

$$P_E(E) = \hat{E}\, e^{-E/\hat{E}}$$
(7)

where $\hat{E}$ = average kinetic energy. The fraction that flies far enough to escape the outer cone must have $E = mv^2/2 > mgH$ to overcome potential energy, where $m$ and $v$ are mass of a grain and its ejection velocity from the rim of the inner crater. Integrating to find the fraction that escape the outer cone,

$$f_{\mathrm{escape}} = \int_{mgH}^{\infty} \hat{E}\, e^{-E/\hat{E}}\, dE = e^{-mgH/\hat{E}} = e^{-H/a}$$
(8)

which identifies the naturally occurring length scale $a = \hat{E}/mg$. Future work could treat this in more detail, but this explains the form of eq. 5 and hence why the crater depth grows logarithmically.

The form of these equations indicate that the secondary erosion rate increases as $H$ increases, but the primary erosion rate is proportional to $ab$ and thus is constant over time. This is remarkable because it shows that primary erosion rate is unaffected by the mass loading higher in the boundary layer. The avalanching secondary grains are injected into the gas flow in the inner crater with initial rolling kinetic energy that carries them transversely across the inner crater's boundary layer at the inner lip, and they fall into the empty space inside the inner crater before the gas reverses their velocity and carries them out again (fig. 2b). Thus, they take momentum and kinetic energy from the gas over the region of maximum primary erosion near the lip. The constancy of the primary erosion rate despite the mass loading increasing with the secondary erosion rate is a direct contradiction of Robert's erosion equation, which says that mass loading higher in the boundary layer is precisely what determines erosion rate [1].

*3.3. Quantitative Results of Reduced Gravity Experiments*

The third evidence comes from constraining $\langle D \rangle$ directly via reduced gravity experiments. Measuring $\langle D \rangle$ accurately will require experiments with improved methods at more gravity levels, plus measurement of the test material's particle size distribution $P(D)$ in the region of the submicron fines, because the fines contribute significant cohesion to the soil. Measuring these will require protracted efforts with a relatively high level of funding. Here, existing datasets are used to bound $\langle D \rangle$ and $\alpha$. This is done for three reasons. First, it is adequate to reach the major conclusion of the present paper. Second, it responds to the urgency of lunar landing plume effects as lunar campaigns are already being developed. Third, it is a necessary step to identify gaps in the theory and to define measurements that should be made in future experiments, including on the Moon, leading to definition of lunar instruments and tests. That costly effort cannot be justified until we have exhausted what is possible from existing data sets.

Measuring $\langle D \rangle$ is complicated because stability of the slope of the inner crater depends on a balance between gravity and gas traction, and thus, crater morphology changes with gravity, altering the gas flow that follows the contour of the inner crater. This in turn affects the shear stress that the gas applies to the surface, $\tau = \tau(\vec{R})$ in the numerator of eq. 3, so it changes with gravity, $\tau = \tau(\vec{R}|\ g)$, partially offsetting the gravity scaling of the denominator. However, the stability of the inner crater slope at two different gravities for the coarse sand, which has $\alpha = 0$ to excellent approximation, can determine the gravity-scaling of the numerator. This numerator scaling will then be applied to tests at two different gravities again, but with a cohesive material having nonzero $\alpha$, so we can solve for $\langle D \rangle$.

There are no reduced gravity data for materials that have characterized $\alpha$ or $P(D)$ in the ultra-fines range, because it is difficult to measure [32–34], and the importance of this range was previously unknown but will be shown here. Reduced gravity data are available for JSC-Mars-1A, but the closest match to it that has been characterized for ultra-fines is JSC-1A [35,36]. The $\alpha$ for JSC-1A is calculated using the van der Waals force equation [37] parameterized per Götzinger and Peukert [38] and Israelachvili [39], integrated over the particle size distribution for all the grain-to-grain contacts in a volume of soil, yielding $\alpha = 0.43$ to $1.72$ J/m³. This should not be confused with the cohesion from the Mohr-Coulomb relationship, which is usually larger and is a shear parameter involving "apparent cohesion" from the particle shapes [40]. The van der Waals equation scales as $D^2$ while the number of contacts per volume scales as $1/D^3$ (for similar compactivities) so $\alpha$ scales as $1/D$. $P(D)$ for JSC-Mars-1A is shifted by a factor of 4 toward coarser particles, $D_{50} = 80$ μm for JSC-1A and $D_{50} \approx 320$ μm for JSC-Mars-1A, so $\alpha \approx 0.11$ to $0.43$ J/m³ for JSC-Mars-1A.

The analysis here uses tabular data extracted by NASA from videography of reduced gravity flights that had been performed aboard reduced gravity aircraft in May 2008 for 250–300 μm coarse sand and October 2008 for JSC-Mars-1A Martian soil simulant [41]. For the crater depth data, the period of logarithmic growth was isolated and fitted to eq. 4. For each granular material and each gravity level, the fitting parameters $a$ and $b$ were determined and averaged over repeated experiments, enabling an estimate of the errors.

In the following, subscripts are $L$ = lunar gravity test conditions, $M$ = Mars gravity, and $E$ = Earth gravity. Erosion occurs mainly at the upper lip of the inner crater. In that location, the ratio of gas shear stress at any gravity versus at Earth gravity is determined by the force balance condition for stability of soil on the slope at those gravities,

$$\frac{\tau}{\tau_E} = \frac{g(\sin\theta - \mu\cos\theta)}{g_E(\sin\theta_E - \mu\cos\theta_E)} = f_\tau(g) \tag{9}$$

where $\theta = \theta(g)$ and $\mu = \tan\varphi$ is the coefficient of friction of the sand at the free surface, estimated by the angle of repose in the outer crater. $\theta$ and hence $f_\tau$ can be measured from the experiment videos. The change in crater width, $w/w_E = f_w(g)$, can also be measured directly from experiment videos, so $\tau/\tau_E = f_\tau\left(f_w^{-1}(w/w_E)\right)$, telling how the traction of the gas reduces as the crater widens.

Eq. 9 describes stability of a thin layer at the surface, not stability of the bulk material inside a tall wall of soil. For soil slopes that are too high, bulk slumping occurs. Here, the parabolic profiles of the inner craters become extremely steep at gravity levels lower than Mars, so an increase in inner crater width going from Mars to lunar gravity would require a disproportionately larger increase in the inner crater height, and bulk slumping should occur. In the lunar gravity experiments, the crater growth curves displayed vastly greater dispersion than in the higher gravities, indicating stick-slip behavior in the soil, which is characteristic of slumping, and the inner crater widths in lunar gravity were only slightly wider than in Mars gravity despite the vastly different gravities. Lunar gravity cases will be omitted in this step of the analysis since here we are only trying to estimate the shear stress of the gas as a function of crater width, so we need eq. 9 to apply. Also, hyper gravity cases (1.8 g) did not last long enough to achieve the inner/outer crater shape, so eq. 9 does not apply in those cases, and they cannot be used here, either. This leaves only two gravity levels, Mars and terrestrial, to determine $f_\tau$ and $f_w$.

It will take experiments at more gravity levels to directly determine the functional forms of $f_\tau$ and $f_w$. Reduced gravity aircraft typically fly only zero gravity, lunar, Mars, and hypergravity (1.8 g) profiles, and funding levels have never been adequate in this research area to purchase a dedicated flight for intermediate levels. However, the range of inner crater widths as gravity changes was quite small, so errors over the small range will be quite small and the functional form is not crucial to this estimation (and recall we are treating this line of evidence as a clue, not a proof). Rajaratnam and Mazurek [28] showed that the shear stress of a jet impinging on a flat plate decays with radius by a power law, so it is reasonable that $\tau/\tau_E = f_\tau\left(f_w^{-1}(w/w_E)\right) = (w/w_E)^\eta$ here, too. It will be adequate to treat both $f_\tau$ and $f_w$ as power laws, $\tau/\tau_E = (g/g_E)^m$ and $w/w_E = (g/g_E)^n$, so

$$\tau/\tau_E = (w/w_E)^{m/n} \tag{10}$$

The experiments found inner crater slope angles $\theta_M = 66° \pm 5°$ and $\theta_E = 62° \pm 5°$, and the ratio of inner crater widths $(w_M/w_E) = 1.572 \pm 0.089$. These yield $m = 1.26 \pm 0.26$ and $n = 0.467 \pm 0.049$, so $m/n = 2.70 \pm 0.31$. This result compares excellently to Rajaratnam and Mazurek [28] for their test cases with comparable conditions (their test cases 4 and 8), which found a power index between 2.5 and 3.0.

Eq. 10 will now be used to analyze erosion rates in the cohesive JSC-Mars-1A. Since erosion occurs mainly in a narrow strip near the crater lip, the area of erosion scales as the circumference of the inner crater, so the primary erosion rate scales as $abw$, where $w$ is inner crater width. The ratio of erosion rates at two gravities is measurable in the experiments as,

$$F = \frac{(ab)_L \, w_L}{(ab)_E \, w_E} \tag{11}$$

For JSC-Mars-1A this was found to be $F_{\text{JSC}} = 2.37 \pm 0.22$ The ratio of eq. 3 at lunar and terrestrial gravities for the same granular materials and the same gas flow conditions is,

$$F = \frac{\tau_L \, w_L}{\rho_b g_L \beta_L \langle D \rangle + \alpha} \cdot \frac{\rho_b g_E \beta_E \langle D \rangle + \alpha}{\tau_E \, w_E} \tag{12}$$

Equating eqs. 11 and 12, and substituting eq. 10 for the ratio of shear stresses,

$$F_{\text{JSC}} \cong \left(\frac{w_L}{w_E}\right)^{-3.70 \pm 0.31} \frac{\rho_b g_E \beta_E \langle D \rangle + \alpha}{\rho_b g_L \beta_L \langle D \rangle + \alpha} \tag{13}$$

The reduced gravity experiments with JSC-Mars-1A measured $\theta_L = 86°$, $\theta_E = 56°$, $\beta_L = 0.070$, $\beta_E = 0.559$, and $w_L/w_E = 1.33 \pm 0.13$. $\alpha = 0.11$ to $0.43$ J/m³, as calculated above. These yield $\langle D \rangle = 2.5 \, D_{50}$ to $4.6 \, D_{50}$, or $1.1 \, D_{84}$ to $2.0 \, D_{84}$ (1st to 3rd quartile results when using a uniform distribution over the input uncertainties). Thus $\langle D \rangle \approx 1.5 D_{84}$, so the rate-limiting physics is located at only one to two coarse grain diameters above the surface where energy flux in the gas is very small, consistent with the other two lines of evidence,

As a check on the results, the value of $n/m$ can be solved by applying eq. 12 to the coarse sand so $\alpha = 0$. Crater widths cancel from the equation. Using Mars compared to Earth gravity,

$$\frac{(ab)_M}{(ab)_E} = \frac{\tau_M}{g_M \beta_M} \cdot \frac{g_E \beta_E}{\tau_E} \tag{14}$$

with values for $(ab)$ and $\beta$ measured in the experiments finds $\tau_M/\tau_E = 0.250 \pm 0.075$. Using this in eq. 10 with the measured $(w_M/w_E)$ yields $n/m = 3.06^{+0.82}_{-0.59}$. The errors are larger from this method, but the result is within about one standard deviation of the prior result, $m/n = 2.70 \pm 0.31$.

This analysis assumed that $\alpha$ is dominantly the van der Waals energy density in the soil. If $\alpha$ were doubled by the inclusion of electrostatic dipoles on the soil grains, we would still find $\langle D \rangle \approx 3 D_{84}$, deep in the laminar part of the boundary layer. For lunar landings, hot rocket exhaust is a positively charged plasma [42], so it will neutralize electrical dipoles and may create monopoles that repel, resulting in a reduction of $\alpha$. Without electrostatic measurements in both terrestrial and lunar cases, we cannot say more about this, but the experiments indicate a very small $\langle D \rangle$ on the order of $D_{84}$.

## 4. The New Theory: Laminar Sublayer Energy Flux

The weight of the three lines of evidence suggests several things. First, the denominator of eq. 3 is the energy that must be delivered per volume of soil to disassemble it, breaking all the cohesive bonds then lifting the grains only the tiny height $\langle D \rangle \approx 1.5\, D_{84}$. This is presumably the average height a grain must reach before it can be swept downwind by the gas.

Second, the fraction of the gas's energy spent accelerating the grains downwind does not appear in the erosion equation. This implies it is a small fraction of the kinetic energy in the gas, so increasing the mass loading has negligible effect. Hence, the fraction of the energy that is transported from just above the grains downward across $\langle D \rangle$ to induce erosion is insensitive to the mass-loading of the gas above $\langle D \rangle$. This is confirmed by the modeling of Chinnappan et al. [43] which showed that soil grains blown by a lunar lander plume have the same downwind velocity with or without two-way coupling between the grains and the gas; that is, taking energy from the gas via two-way coupling has negligible effect.

Third, because this potential energy at height $\langle D \rangle$ is so tiny, and because energy accelerating the entrained grains downwind after they have been lifted does not play a role, the small quantity $\alpha$ is significant. The rate of erosion beneath a lunar lander should therefore be controlled by the ultra-fines in the soil, because they overwhelmingly dominate $\alpha$ (as shown in the companion paper, "Erosion rate of lunar soil under a landing rocket, part 2: benchmarking and predictions," *this issue*). These particles constitute only 1% of the mass of the soil so were considered negligible in prior work and remain poorly characterized for both the real and simulated lunar soils. It seems counter-intuitive that such a small mass fraction of the soil would control the soil's response to rocket exhaust, but this is consistent with the observation that erosion is a low-energy process. If $\langle D \rangle$ is such a small value, placing the rate-limiting physics at the very bottom of the laminar sublayer where energy transport is small, then it is indeed low-energy.

Fourth, the numerator must be the effective rate of kinetic energy delivered by the gas downward across the height $\langle D \rangle \approx 1.5\, D_{84}$. In the rarefied lunar case, the flow in the entire boundary layer should be laminar. Rocket exhaust leaving the nozzle is inviscid then comes to a stop in the stagnation region, so the Reynolds number is small; then it transitions toward free molecular flow as the gas moves radially away from centerline expanding into vacuum. The lengthening mean free path lengths of the gas molecules truncates the potential turbulent spectrum at ever larger eddy sizes, suppressing the triggering and growth of turbulence. Even if some turbulence develops, the flow near the surface is laminar. The Roughness Reynolds Number for a soil with broad particle size distribution may be approximated [8] as

$$\text{Re}_\theta \cong \frac{3.5 D_{50}\, \rho_0 v_*}{\mu} \tag{15}$$

where $\rho_0$ = gas density near the surface, $v_* = \sqrt{\tau/\rho_0}$ is the friction velocity, and $\mu$ = dynamic viscosity of the gas. $\text{Re}_\theta < {\sim}4$ indicates hydraulically smooth flow whereas $\text{Re}_\theta > {\sim}60$ indicates flully rough flow. Using Roberts' plume flow equations to calculate conditions during an Apollo landing, $\text{Re}_\theta$ varies from 0.005 to 0.1, so the flow is hydraulically smooth. Therefore, a laminar sublayer will exist. The thickness of the laminar sublayer is $\delta \cong 5\mu/\sqrt{\rho_0 \tau}$ [8], which calculates to ~10 cm, far thicker than $\langle D \rangle$. Transport of momentum and energy across the height $\langle D \rangle$, deep in the laminar sublayer, can therefore be only by thermal molecular diffusion, which scales by the molecular thermal velocity, $\overline{v_T}$.

For the terrestrial experiments, Phares et al.[44] found that radial flow from a circular jet (with a similar range of jet Reynolds numbers) impinging on a plate will transition from laminar to turbulent at radial distance $r \cong 3 H_e$ where $H_e$ = height of the jet exit plane. Here, the inner craters were typically no wider than $r = 0.2 H_e$ to $0.4 H_e$, well less than the turbulent transition. Thus, both terrestrial experiments and the lunar case had laminar boundary layers.

Defining $\rho_0$ and $v_0$ as the local gas density and velocity at or just above $\langle D \rangle$, for the lunar case, the shear stress just above $\langle D \rangle$ per molecular kinetics is $\tau = \rho_0 v_0 \overline{v_T}/6$, and the downward energy flux across that height is $E = v_0 \tau/2 = \rho_0 v_0^2\, \overline{v_T}/12$. Comparing with eq. 3 identifies $C$ with $v_0$ multiplied by no more than a proportionality constant, $\varepsilon$, since all other variables have been accounted for. Because the numerator of eq. 3 is an energy flux and the denominator is the energy consumed eroding soil, $\varepsilon$ indicates the fraction of the influx of energy that is converted to mechanical work lifting grains. As such it will be called the erosion efficiency.

The local erosion rate (kg/m²/s) for both terrestrial and lunar atmospheres can now be written,

$$\dot{m} = \rho_b \frac{\varepsilon \left(\frac{1}{12}\rho_0 v_0^2\, \overline{v_T} - E_{\text{th}}\right)}{\rho_b g \langle D \rangle + \alpha}, \qquad E > E_{\text{th}} \tag{16}$$

or $\dot{m} = 0$ if $E < E_{\text{th}}$, where $E_{\text{th}}$ = threshold value of $E$ capable of initiating erosion [45,46]. $E_{\text{th}}$ had been omitted from eq. 3 since the terrestrial experiments were far above threshold and it was not identified in the results, but in the lunar case the lander descends until erosion first begins so this threshold is important. $\beta = 1$ for a flat slope on the Moon so it was omitted. With this choice of $\langle D \rangle$, $\alpha$ is affirmed to be the cohesive energy density. $\varepsilon$ must be no more than weakly dependent on the parameters already included in eqs. 1 and 2 for consistency with the experiments, which constrains its interpretation. Details of the erosion process, such as different sized particles lifting at different rates and some particles being redeposited in the erosion zone, as described in section 3.1, may affect the net erosion rate so in this model these effects are collected into $\varepsilon$. The companion paper will test this equation and determine the value of $\varepsilon$.

5. **Summary**

The erosion process can be summarized conceptually as follows. Gas flows over the surface at high velocity. Most of that energy advects away without doing anything to the soil. Molecular diffusion in the gas flow causes a tiny fraction of this kinetic energy to cross the plane $\langle D \rangle \approx 1.5\ D_{84}$, creating stagnation pressure below the roughness height of the static grains. This puts a net force under the grains to lift them. Most of this compressed gas energy below $\langle D \rangle$ diffuses back into the horizontal flow above $\langle D \rangle$ and advects away without doing mechanical work on the grains. Some fraction of the stored energy, $\varepsilon$, lifts the grains to the height $\langle D \rangle$ where they are exposed to the kinetic energy of the horizontal gas flow and are accelerated away. The energy they consume while being accelerated is a negligible fraction of the high energy in the flow above $\langle D \rangle$, so the entrained mass loading has negligible effect on the erosion rate below $\langle D \rangle$. If the diffusion of energy below $\langle D \rangle$ is too small, the stagnation pressure will be inadequate to move the grains so the gas molecules will simply diffuse back out above $\langle D \rangle$ and be advected away without doing any mechanical work. Just below the threshold of erosion, grains might be lifted slightly or jostled, but they fall would back to the surface locally without knocking others free. At slightly higher energy flux the induced grain dynamics become adequate to overcome the potential energy barrier to clear the neighboring grains, $\langle D \rangle$. This predicts a threshold condition for erosion, $E_{\text{th}}$.

These results were based upon clues in the existing data sets. Full confirmation of the results will require expensive experimental campaigns using improved methodology, including measurements at multiple gravity levels and instruments flown to the lunar surface. For now, this result will be partially tested in the companion paper, "Erosion rate of lunar soil under a landing rocket, part 2: benchmarking and predictions" (*this issue*), which compares the predictions of the new theory to the observed optical density of the blowing dust during an Apollo landing.

6. **Conclusions**

Erosion of soil in lunar landings, where saltation plays no role, is a low energy process. Roberts' hypothesis for lunar soil erosion, that it is a shearing process linear with gas shear stress and governed both by shear strength of the soil and by acceleration of the eroded particles reducing shear stress of the gas, is incorrect and should not be used. Instead, erosion is an energy flux phenomenon, scaled by molecular diffusion in the laminar gas flow just above the tops of the stationary grains. It is resisted by the potential energy to lift each grain to just more than the average height of the neighboring sand grains (after which it is directly exposed to the horizontal gas flow), and by the cohesive energy density of the soil. These are small quantities of energy, corresponding to the small energy flux in the laminar gas flow at the bottom of the boundary layer. Cohesive energy density is supplied to the soil mainly by the ultra-fine particles smaller than 3 $\mu$m, which are poorly characterized. Existing reduced gravity data sets on soil erosion were obtained before these factors were known, so the experimental methods had problems.

Only a rough analysis was possible here from these existing data sets. Future experiments should carefully calibrate the ultra-fines content of the soils, develop techniques to quantify the cohesion and detect bulk slumping in the soil, and operate in multiple gravity levels between terrestrial and lunar gravity. Experiments are also needed (and are currently in-work) that control the temperature of the gas independently to verify the role of molecular thermal velocity. The new erosion equation is used in the companion paper to calibrate the erosion efficiency parameter via the optical density of blowing soil in the Apollo landings, and it shows that erosion rate is significantly higher than previously believed. Continuing this research should result in better international policy for the blast effects of lunar landings.

**Funding**

This work was supported by NASA grant numbers 80NSSC20K0810 and NNA14AB05A.

**Author Contributions**

Conceptualization, formal analysis, writing – original draft, and writing – review and editing: PM

**Competing Interests**

The author declares there are no competing interests.

**Data and Materials Availability**

Tabular data for figures 1c and 1d are in the Supplemental Material. Reduced gravity experiment videos and extracted tabular data are available at https://doi.org/10.5281/zenodo.10611667.